# SSCU: an R/Bioconductor package for analyzing selective profile in synonymous codon usage

Yu Sun[1,2,3,*] and Siv G.E. Andersson[2]

[1]Guangdong Provincial Key Laboratory of Protein Function and Regulation in Agricultural Organisms, College of Life Sciences, South China Agricultural University, Tianhe District, Guangzhou, Guangdong 510642, PR China. [2]Key Laboratory of Zoonosis of Ministry of Agriculture, South China Agricultural University, Guangzhou, Guangdong 510642, PR China. [3]Department of Molecular Evolution, Cell and Molecular Biology, Science for Life Laboratory, Uppsala University, 752 36 Uppsala, Sweden

*To whom correspondence should be addressed.

# Abstract

**Background**

Synonymous codon choice is mainly affected by mutation and selection. For the majority of genes within a genome, mutational pressure is the major driving force, but selective strength can be strong and dominant for specific set of genes or codons. More specifically, the selective strength on translational efficiency and accuracy increases with the gene's expression level. Many statistical approaches have been developed to evaluate and quantify the selective profile in codon usage, including *S* index and Akashi's test, but no program or pipeline has been developed that includes these tests and automates the calculation.

**Results**

In this study, we release an R package SSCU (selective strength for codon usage, v2.4.0), which includes tools for codon usage analyses. The package identifies optimal codons using two approaches (comparative and correlative methods), implements well-established statistics for detecting codon selection, such as *S* index, Akashi´s test, and estimates standard genomic statistics, such as genomic GC3, RSCU and Nc.

**Conclusions**

The package is useful for researchers working on the codon usage analysis, and thus has general interest to the biological research community. The package is deposited and curated at the Bioconductor site, and has currently been downloaded for more than 2000 times and ranked as top 50% packages.

**Keywords**

Codon usage, Selection, R, Akashi's test, *S* index

**Background**

The central dogma is the foundation of modern molecular biology; it describes the information processing flow within a biological system in the direction from DNA to RNA to protein. An important step in the process is to decode the information stored in the cryptic nucleotides sequences into functional protein molecules. Sixty-four tri-nucleotide units (codons) code for twenty amino acids, and thus most of the amino acids have more than one matching codon. The codon choice for an amino acid is not random; mutation and selection are the two major forces shaping the pattern. For most of the genes, mutation is the major force determines the codon usage, but selective forces become increasingly important with the genes' expression level due to selection for translational efficiency and accuracy [1, 2].

Currently, the most frequently used program is CodonW, which was developed 17 years ago. It calculates several codon usage metrics such as the overall frequency of codons, GC3s values, RSCU (relative synonymous codon usage) and Nc values (effective number of codons) [3]. Codon adaptation index (CAI) is another widespread technique for codon usage bias, which is included in a R pack 'seqinr'. Other programs, such as INCA, OPTIMIZER and CAIcal includes functions to calculated codon indices such as effective number of codons (NC') and CAI [4-6]. Except the well-established statistics, several theoretical and mathematical methods have been developed to detect the influence of selective forces, including Sharp's $S$ index and Akashi's test [2, 7]. The paper for the $S$ index algorithms has been cited for more than one-hundred times, but no program has been developed to automate the calculation yet. The Akashi's test is a population approaches to evaluate the strength of translation selection in codon bias, and it has been suggested that the Mantel-Haenszel test in the open-source statistical package R can do the calculation. However, we showed that the test in R is not appropriate since it does not perform the right calculation [8], thus the tool is still missing for the analysis.

Except the statistical analysis for codon selection, identifying codon set under selection are also one of the central task for codon studies. Various approaches have also been developed to identify the optimal codons, which mainly categorized as comparative and correlative methods [7, 9, 10]. The comparative method focusses on the highly expressed (mainly ribosomal genes) and compare them with the lowly expressed gene sets. Rather, the correlative method test the correlation on all gene set to identify the most the biased codon [9]. It has been debated for which approaches is better for the optimal codon identification, but no final conclusion has been drawn yet [11, 12]. Thus, tools for the optimal codon identification are needed to solve the sophisticated problem.

In this study, we developed an R package SSCU (Selective Strength for Codon Usage) for codon analyses, including functions to calculate the standard $S$ index, Akashi's test, optimal codon identification, low frequency optimal codon identification, RSCU (relative synonymous codon usage) value [13], and general genomic statistics, such as GC3. This package is useful for biological researchers interested in the codon usage pattern, and it is a good complement to CodonW. The package is deposited in the curated bioinformatics tools site Bioconductor. Although only recently released, the package has been downloaded for more than times, and ranked as the top 50% download among all packages (as in Nov 1, 2017).

**Implementations**

The package is written in programming language R, and curated in the bio-medical and bioinformatics program sites 'Bioconductor'. The current version of SSCU 2.4.0, which has 8 functions, and the brief overview is introduced below:

1 Overall mutation and selection pattern

To estimate the overall mutational and selective patterns in codon usage, the SSCU package provides 2 functions: *s_index* and *genomic_gc3*. The *s_index* calculates the S index [7] to quantify the strength of translational selection in the highly expressed genes. The S index is comparable among species; thus it is very useful in detecting relative translational selection on codon usage in different genomes. The function *genomic_gc3* calculates the overall genomic GC3 content, and can be used to infer the strength of the mutation pressure. The function concatenates all the CDS sequences into one long sequence string and calculates the overall genomic GC contents at third codon positions.

2 Optimal codons identification and statistics calculation

Several methods have been developed to identify "optimal codons", which are defined as the codons favored by translational selection. Generally, the methods can be categorized into two groups: comparative methods and correlative methods. The comparative method focuses on the highly expressed genes and identifies a list of codons that are statistically more abundant in this genes set compare to the reference sets [10]. The function *op_highly* in the package performs this calculation. The correlative method, one the other hand, calculates the frequency of each codon in all genes, and predicts the optimal codons by correlation test [9]. To conduct the test, Nc or Nc' (effective number of codons) values need to be input to the function [14, 15]. Nc is one of the most population index to study the level of codon usage bias, and the calculation of Nc is similar to the effective population size in the population genetics study. In this SSCU package, the Nc value can be calculated by the *op_highly_stats* function. Nc' is an new version of Nc, and it can be computed by ENCprime or INCA program [15]. The Nc or Nc' data can be pipe into the function *op_cor_CodonW* or *op_cor_NCprime* to calculate the optimal codons by the correlative method.

The function *op_highly_stats* calculates detailed codon usage statistics, including the RSCU value [16], the total number of codons and Nc for both the highly expressed and the reference set of genes. It also contains optimal codon list inferred by the comparative method. Low frequency codons are sometimes identified as the optimal codons [8]. These codons can be found by the *op_highly_stats* function with manually inspection, but it is easier to do so with the function *low_frequency_op*, which only output the codon list and statistics for the low frequency optimal codons.

3 Akashi's test

The Akashi's test evaluates the selection on translational accuracy for the coding sequences [2]. The theory proposes that optimal codons generally correspond to abundant tRNAs, and are translated more accurately compared to the non-optimal codons. Thus it is selectively preferred to use the optimal synonymous codons in the evolutionary conserved and important amino acid sites. The test calculates the association between optimal codons and amino acids, by comparing the codon constitution between targeted and orthologous reference sequence. To do it, the first step is to tabulate the codons as shown for each amino acid in each gene (Table 1). Thus for each gene, 18 contingency table should be available as in Table 1, excluding codons for Methionine and Tryptophan. Then the test summarizes data in all the contingency table for a given gene set. In the package, we attached a Perl script *make_contingency_table.pl* to count and tabulate the conserved and variable sites for sequence alignment file, then using function *akashi_test* for the calculation. The algorithms for Akashi's test can be referred to the Akashi's paper [2], and detailed usage of all the functions can be referred to SSCU Bioconductor site or supplementary file 1.

Table 1. Codon number tabulation for Akashi's test.

| Certain amino acid | Conserved | Variable |
|---|---|---|
| Optimal | a* | b* |
| Non-optimal | c* | d* |

* a is the number of optimal codon coding for conserved amino acid

b is the number of optimal codon coding for variable amino acid

c is the number of non-optimal codon coding for conserved amino acid

d is the number of non-optimal codon coding for variable amino acid

**Results**

To illustrate the function usage of the SSCU package, we analyzed the genomic and codon usage features for a few *Lactobacillus* genomes. Three *Lactobacillus* genomes, including *Lactobacillus fermentum* IFO 3956 (accession number NC_010610), *Lactobacillus sakei* subsp. sakei 23K (NC_007576) and *Lactobacillus mellis* Hon2 (JXBZ01000000) were downloaded from GenBank, NCBI, and the *S* value, GC3s and optimal codon lists for each genome computed by *s_index*, *genomic_gc3*, *op_highly* function respectively (Figure 1). Two out of the three *Lactobacillus* species has genomic GC3 values of lower than 40%, which is the general mutational pattern for most *Lactobacillus* species [8]. However, one species *L. fermentum* displays a strong mutational shift manifested in a genomic GC3 value of >60%. For the translational selection, the three species also show the whole spectrum of *S*-values, including one species *Lactobacillus mellis* without detectable translational selection (*S* index < 0.2). Thus shift in both mutational and selective pressure in codon usage has been observed in the *Lactobacillus* species, which is an interesting pattern for species that only recently diverged from a common ancestor.

Various approaches have been developed for optimal codon identification, which can be mainly divided into two categories: comparative methods and correspondence methods. The two method generally predict similar sets of optimal codons for a given species, such as in *L. sakei* (Figure 1). But for some species, the methods computed rather different sets, such as in *L. fermentum* (Figure 1), that the comparative method identified mainly AU-ending codons as optimal codons, whereas the correlative method identified GC-ending codons. There is an ongoing debate about whether one approaches is superior than another, but no conclusion has been drawn yet [11, 12]. Sun et al suggests that the GC content shift is the major reason for the variable prediction by the two approaches [8]. Thus further studies are needed to disentangle different factors and validate or find the limitation for each approach in the optimal codon identification.

Except the previous functions, the package also includes functions to calculate Akashi's test, which is a population approach to evaluate the strength of codon selection on translation accuracy, RSCU value, which estimates the relative usage of codons, number of observed codons and number of expected codons for each amino acid. Part of the results were shown in Figure 2.

**Conclusions**

The SSCU package will be useful for the biological community who are interested in conducting codon usage analysis and disentangle the selective and mutational forces on the genomic level. The package includes functions to perform the *S* index, Akashi's test and to identify optimal codons by the correlative method, and it is complementary to the widely used CodonW program. Although it has only been available for a few months, the package has already by downloaded for more than 2000 times in the Bioconductor depository site.

**Figures**

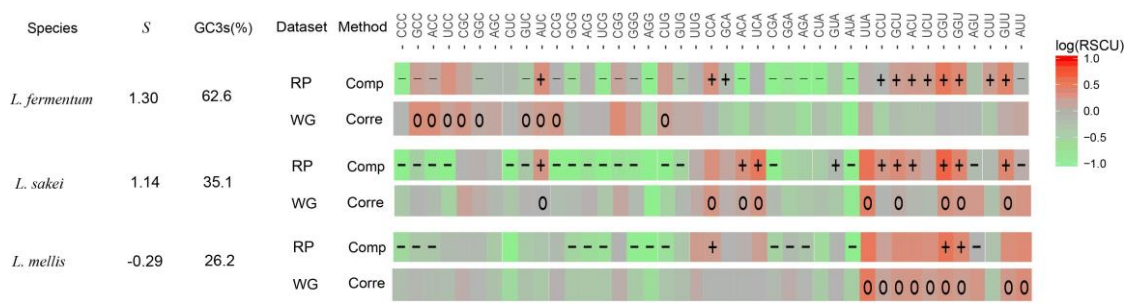

**Fig 1. S, GC3s value and optimal codons identified by different methods in three *Lactobacillus* species.** The color in the panel shows the log(RSCU) value for the ribosomal proteins (RP) and whole genome (WG). The symbol "+" represents optimal codons detected by the comparative method, and "o" represent optimal codons by the correlative method based on Nc values. The data for the figure was took from the Sun Y et al (2016).

```
codon aa rscu_high rscu_ref high_No_codon high_expect_No_codon ref_No_codon ref_expect_No_codon p_value symbol
TTT   F    0.66      1.10          58              88                10540         9539          0.003    -
TTC   F    1.34      0.90         118              88                 8538         9539          0.005    +
TTA   L    2.58      2.54         198              77                16254         6399          0.895    NA
TTG   L    0.56      1.09          43              77                 6971         6399          0.001    -
TCT   S    1.10      1.04          64              58                 5263         5062          0.789    NA
TCC   S    0.12      0.38           7              58                 1937         5062          0.001    -
```

**Figure 2. The detailed codon usage statistics output by the *op_highly_stats* function in the SSCU package.**

**Additional files**

Supplementary File 1. The user manual for SSCU package.

**Availability and requirements**

  Project name: SSCU

  Project home page: https://bioconductor.org/packages/release/bioc/html/sscu.html

  Operating system(s): Linux

  Programming language: R

  Other requirements: R 3.3 or higher, Tomcat 4.0 or higher

  License: GPL > 2.0

  Any restrictions to use by non-academics: license needed

**Declarations**

**Ethics approval and consent to participate**

Not applicable

**Consent for publication**

Not applicable

**Competing interests**

The authors declare that they have no competing interests

**Authors' contributions**

YS and SGE initiated the project. YS coded the package and analyzed the data. YS and SGE wrote the paper.